# A Proposed Algorithm to improve security & Efficiency of SSL-TLS servers using Batch RSA decryption

R.K.Pateriya, J.L.Rana, S.C. Shrivastava, Jaideep Patel
Department of Computer Science & Engineering
Maulana Azad National Institute of Technology
Bhopal, India
Emails: rkpateriya@manit.ac.in, jl_rana@yahoo.com, scs_manit@yahoo.com, jaideep1983@gmail.com

*Abstract* — Today, Internet becomes the essential part of our lives. Over 90% of the e-commerce is developed on the Internet. A security algorithm became very necessary for producer-client transactions assurance and the financial applications safety (credit cards, etc.) The RSA algorithm applicability derives from algorithm properties like: confidentiality, safe authentication, data safety and integrity on the internet. Thus, this kind of networks can have a more easy utilization by practical accessing from short, medium, even long distance and from different public places (Internet Cafe, airports, banks, commercial centers, educational institutes, etc.) the immensity of resources offered by internet. RSA encryption in the client side is relatively cheap, whereas, the corresponding decryption in the server side is expensive because its private exponent is much larger. Thus SSL/TLS servers become swamped to perform public key decryption operations when the simultaneous requests increase quickly. The Batch RSA method is useful for such highly loaded web server. In our proposed algorithm by reducing the response time & client's tolerable waiting time an improvement in performance of SSL-TLS servers can be done. The proposed algorithm should provide the reasonable response time and optimizes server performance significantly. At Encryption side, to withstand many attacks like brute force attack, subtle attack etc. we also adapted a parameter generation method, which sieve all the parameters strictly, and filter out every insecure parameter.

## I. INTRODUCTION

SSL/TLS handshake protocol is a typical key encapsulation approach for secure communication, implementation of RSA algorithm in SSL/TLS is computationally imbalanced for client and server. As a result RSA encryption in the client side is relatively cheap, where as the corresponding decryption in server side is expensive because its private exponent is much larger. In previous batch method some flaws were there which can be improved with proposed algorithm. Batching parameter is optimized when integrating user's requirements for Internet Quality of Service (QoS). To select the optimal batching parameters, not only the server's performance but also the client's tolerable waiting time is considered. Based on the analysis of the mean queue time, batching service time and the stability of the system, a novel batch optimal scheduling algorithm which is deployed in a batching Web server is proposed. In addition the proposed scheme in this paper models the minibatching into existing algorithm proposed in [4].

The security of RSA algorithm is based on the difficulty of factoring large numbers which is almost impossible for 1024 bit numbers. To be able to generate the RSA parameters, one has to decide on the maximum allowed length of each of these parameters. This will be reflected on the security of the system. Initially the prime factors p and q number should be chosen to generate the modules number n. This n should be of a certain length, which is controlled by the generating number (n-bit). Varying this length will successively change the length of both p and q since they are the factors of n. so we adapted parameter generation method to make the secure transmission more effective and to withstand different attacks like

(1) Brute Force attack
(2) Subtle attack    etc.

### A. Preliminaries

Batch RSA Decryption

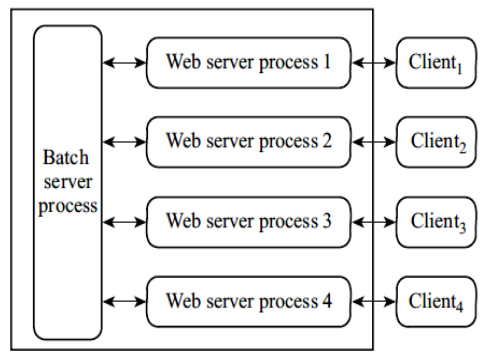

Fig 1 Batch decryption server

For b different ciphertexts $c_1, c_2 \ldots c_b$ encrypted with the different public key ($e_1, e_2 \ldots e_b$). Our goal of the batch decryption is to get the correspondent plaintexts $m_1, m_2, \ldots, m_b$ via one decryption operation. It is well known that decryption operation is computational intensive due to the modular multiplication operation. By using a batch RSA decryption, we aim to efficiently improve the decryption efficiency.

System architecture in Fig1 consists of two kinds of processes the batch server process and the web server process. The batch server process schedules and performs batch RSA decryptions. The web server processes perform SSL/TLS handshake with each client and send decryption request to the batch server. The batch server uses a round robin strategy to aggregate the decryption request and complete batch decryption. The decryption results are returned to web server processes which interact with SSL/TSL handshake clients [9].

Batching of client request has two advantages. First it improves the throughput of an RSA. Second batching significantly improves the behavior of a system if the arrival of message is bursty. Batching is effective during peak time in which more messages arrive than system can handle.

## II. PROPOSED OPTIMAL BATCH SCHEDULING ALGORITHM

In previous algorithm [4] some flaws were there, if no. of customers in batch are not completely filled then server has to wait for request so that batch of queues is completely filled with client request arrived for decryption, then server do batch decryption, but it increase the client's tolerable waiting time of client and response time of server.

In our proposed scheme we perform minibatching in batch of queues in algorithm from[4], made for clients request decryption so that requests which are waiting for long time can be decrypted with the web server, so tolerable waiting time & response time can be decreased .so it improves the performance of web server. Value of $T_b$ Taken from [4].

$$T_b = \frac{(3n^3 + n^2 (42b + k(3b^3 + 3b) - 1))bT_{rsa}}{b(3n^3 + n^2)}$$

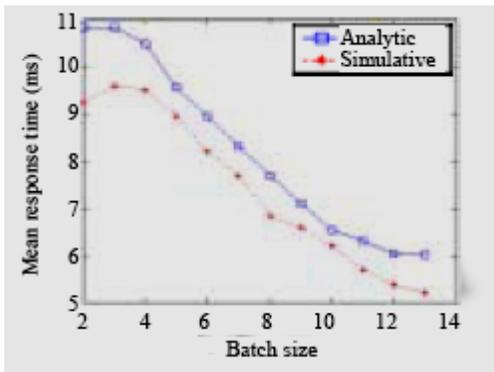

Fig 2 Mean response time speedup non-batching Scheme against non- batching

Figure 2 shows [4] the comparison of the mean response time of the batching schemes with the non-batching scheme. The vertical axis is the mean response time over batch size divided by the mean response time with non-batching scheme. Performance of server can be improved with the proposed algorithm.

Proposed optimal batch scheduling algorithm

---

Step1: Find out the solution of b

Input: $T_t, \lambda$;

Output: $T_b$, optimal batch size

Begin

Compute the max batch size. maxbatchsize= Int($0.4\lambda T_t+1$)

If (maxbatchsize<=1) then do

Conventional_RSA_decryption(); return;

For (b=2; b<=maxbatchsize;b++)

{Success=false;

Compute $T_b$;

If ($T_b$<b/$\lambda$) then {Optimalbatch size=b;

Success=true ;}}

If (! success) then return;

If (success) then goto step2 ();

End

Step2:Optimal batch scheduling

Input: optimal batch size, $T_t$;

Output :optimal batch scheduling.

Begin

Construct b queues (Q ($e_1$), Q ($e_2$)…Q ($e_b$))

For every exponent $e_i$ maxtimer=0;

Assign the exponents { $e_1,e_2,e_3….e_b$) to different clients

using round robin strategy to server hello message

While(Q($e_1$)!=Null or ($e_2$)!=Null..or ($e_b$)! =null)}

If client arrived then (match client. exponent=ei

Enqueue(Q($e_i$).client ); initialize Client timer}

If (Q($e_1$)!=Null and Q($e_2$)!=Null….and Q($e_b$)!=Null)

Then (do batch decryption () ;

Reset server_waiting time;

Update queues (Q($e_1$),Q($e_2$),…..Q($e_b$));

Else {  for ( j=1;j<=b;j++)

{

    If ((Q ($e_j$!=Null) and Q($e_j$).head timer>=maxtimer))

```
{ maxtimer={Q(ej).head.timer);}}
If (server waiting time >=Tt-maxtimer} then
{ do batch_decryption();
Reset server_waiting _time;
Update queues
Else
For j=1 to b
If any (Q (ej)=null) then
 If Q(ej).head.timer>=maxtimer then
Make miniBatch of queues Q (ej)
Apply batch_decryption ();
Update Queues;
Repeat until batch size =1;
If batch size=1;
Do conventional_RSA_decryption;
End
```

## III. RSA PARAMETERS

The RSA algorithm has some important parameters affecting its level of security. It consists of some high order mathematical operations performed on some parameters in a certain order. These parameters control the level of security of the encrypted data. It was shown [6] that complexity of decomposing the modules into its factors is a function of the modules length itself. The importance of this length is also reflected on the security of the public key making it more difficult to be detected.

The importance and effect of changing the RSA Parameters are analyzed such that one parameter is changed at a certain time and the others are kept fixed.

It is conjectured that if n is generated by picking at random two big primes and multiplying them, then factoring n is an intractable problem. Also computing d given e and n is as hard as factoring n. This is the assumption of the RSA; clearly if factoring is easy then RSA assumption fails. The RSA algorithm provides excellent protection of voice and data.

**Changing the modules length:** Changing the maximum length of the generating number n-bit to generate the modules n will affect the other parameters as shown in Table I

TABLE I. THE EFFECT OF CHANGING THE MODULES NUMBER

| n-bit | p length | q length | n length | e length | d length | C length |
|---|---|---|---|---|---|---|
| 500 | 76 | 76 | 151 | 21 | 151 | 151 |
| 600 | 91 | 91 | 181 | 21 | 181 | 181 |
| 700 | 106 | 106 | 212 | 21 | 211 | 211 |
| 800 | 121 | 121 | 242 | 20 | 241 | 241 |
| 900 | 136 | 136 | 272 | 20 | 272 | 271 |
| 1000 | 151 | 151 | 302 | 20 | 301 | 301 |
| 1024 | 155 | 155 | 309 | 20 | 308 | 309 |
| 1200 | 181 | 181 | 362 | 21 | 361 | 362 |
| 1500 | 227 | 227 | 453 | 21 | 452 | 452 |

It is clear that increasing the maximum limit on the length of the modules number will increase the length of both p and q factors. The length of the secret key d and the length of the encrypted message c are also increased at the same rate as illustrated in Fig. 1.

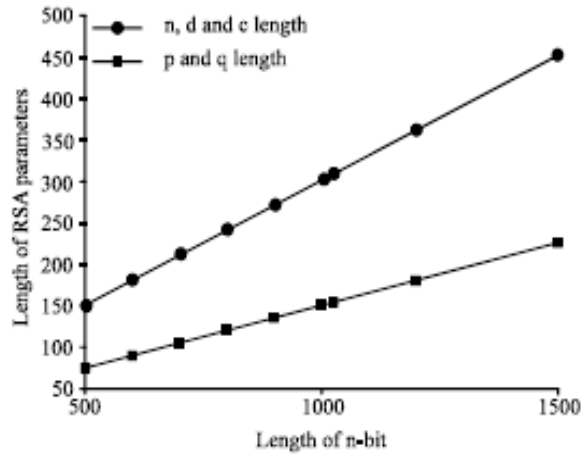

Fig 3    Modules length vs. RSA parameters length

Increasing the n-bit length will provide a more secure value for the private key d, since larger d means more security where as the public key e does not have that importance here

## IV. KNOWN ATTACKS TO RSA ALGORITHM

Methods for attacking RSA algorithm could be classified into two kinds:

Brute force attack. This kind of attack doesn't care any special parameters.

(1) Exhaustive attack: This kind of attack traverses all possible values for d, or tries all possible combinations for 1s in d till the attacker finds the correct d.

(2) Factorization attack: This kind of attack factorizes the module of RSA and gets p and q. The widely used methods at

present are quadratic sieve, generalized number field sieve, and special number field sieve.

Subtle attack. This kind of attack aims at the mathematic feature of some parameters.

(1) Attack by multiplication of small primes: This kind of attack makes use of that there is no big prime factor in p+1 or p-1 or q+1 or q-1. By calculating the modular power with the multiplication of a chain of small primes as the exponent, one gets p and q subtly.

(2) Square attack: When p and q are too close to each other, this kind of attack works. By computing n , one can easily get the real p and q via repeated tests.

(3) Iteration attack: This kind of attack repeats the modular power calculation in the encryption procedure again and again to recover the plain text.

(4) Low private exponent attack: This kind of attack makes use of continued fraction to get approximate plain text. It includes Wiener attack and Boneh-Durfee attack.

By using parameter generation algorithm from [3] so it can protect our message from different attacks.

iteration attack. Length (d)>=length (φ (n))*0.292 can resist low private exponent attack (Wiener attack requires that length (d)>= length (φ (n))*0.25, while Boneh-Durfee attack requires that length (d)>=length (φ (n))*0.292). It is enough for length (d) to be bigger than 80 in order to withstand exhaust attack, for 280 rounds exhaust is complex enough. In order to withstand combination attack, the Hamming weight of d should not be too large or too small. If there are too many or too few 1s in d, one could exhaust every possible position for 0 or 1, and then get the correct d. Hence, according to the bit-length of d, we should guarantee that the Hamming weight of d could make the number of combinations more than 280 (more complicated than exhaust attack).

## V. CONCLUSION & FUTURE WORK

This paper proposed a framework for Batch RSA decryption technique improvement. We applied concept of minibatching in our algorithm due to that Batch RSA decryption time can be reduced to some extent.

The advantage is that proposed algorithm reduce the mean response time, client's tolerable waiting time so that efficiency is improved. By adapted parameter generation algorithm we can protect our data from many security attacks on the internet. Our future work is to implement this proposed algorithm.

ACKNOWLEDGMENT

The Success of this research work would have been uncertain without the help and guidance of a dedicated group of people in our institute MANIT Bhopal. We would like to express our true and sincere acknowledgements as the appreciation for their contributions, encouragement and support. The researchers also wish to express gratitude and warmest appreciation to people, who, in any way have contributed and inspired the researchers.

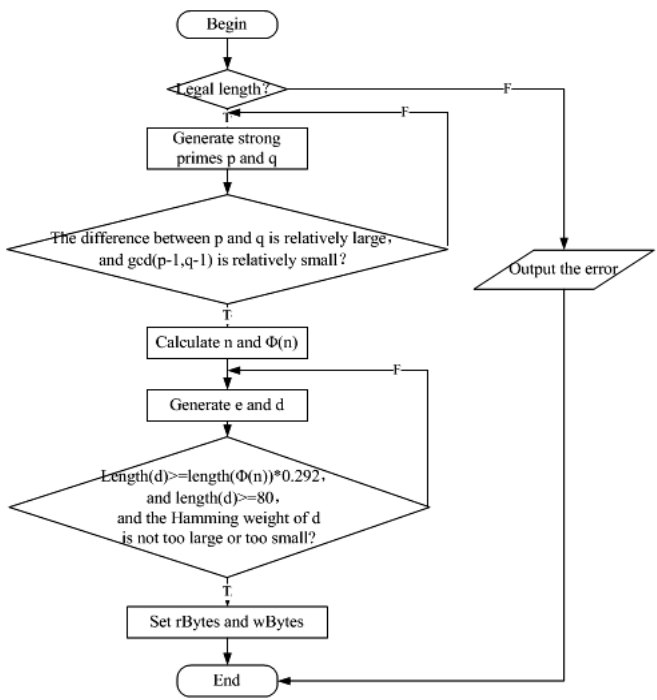

Fig 4 Parameter generation algorithm

Here, p and q are strong primes which can withstand the attack by multiplication of small primes. The difference between p and q is relatively large, which can withstand the square attack. That gcd (p-1,q-1) is relatively small can avoid